\documentstyle[aps,preprint]{revtex}
\begin{document}
\draft
\title{Dynamics of Small Spin Polaron in the Three-Band Model of
Two-Dimensional
Spherically Symmetric Antiferromagnet.}
\author{A.F. Barabanov}
\address{Institute for High Pressure Physics, \\
Troitsk, Moscow region, 142092, Russia.}
\author{R.O. Kuzian}
\address{Institute for Materials Science,\\
Krjijanovskogo 3, Kiev, 252180, Ukraine}
\author{L.A. Maksimov}
\address{Russian Research Center, Kurchatov Institute, \\
Kurchatov sq.46, Moscow, 123182, Russia.}
\author{E.\u Z\c asinas}
\address{Institute for High Pressure Physics, \\
Troitsk, Moscow region, 142092, Russia.}
\date{}
\maketitle

\begin{abstract}
The retarded Green's function $G({\bf k},\omega )$ of a 
single small spin polaron in the three-band model for the 
CuO$_2$ plane is calculated in the self-consistent Born 
approximation. It is shown that such a spin polaron is a good 
quasiparticle excitation for realistic values of 
spin exchange $J$ and effective hopping $\tau $. The polaron 
spectral density $ A_p({\bf k},\omega )$ demonstrates small 
damping in contrast to the results of calculations starting from 
the bare hole, i.e. the pole strength of the energetically 
low-lying quasiparticle peak $Z_p( {\bf k)}$ varies from 50\% 
to 82\% for $J/\tau \propto 0.1\div 0.7$.  The quasiparticle peak 
dispersion reproduces the main features of the bare polaron 
spectrum $\Omega _k$ near the band bottom.

The spherically symmetric approach is used for the description of 
spin excitations. It makes it possible to consider the quantum 
antiferromagnetic background without the spontaneous symmetry 
breaking and the unit cell doubling.

The new method of the self-consistent calculation, based on 
continued fraction expansion of Green's function, is represented 
in details.  The method preserves the proper analytical 
properties of the Green's function and provides the possibility 
to analyze the nature of its singularities.  \end{abstract}



\section{Introduction}

Up to date much theoretical work has been devoted to the problem of a hole
motion in two dimensional (2D) s=1/2 quantum antiferromagnet (AFM) \cite
{dago}. The important question is whether a hole injected in the undoped
ground state behaves like a quasiparticle (QP). This problem is mainly
investigated within the framework of self-consistent Born approximation
(SCBA) for the $t-J$ model \cite{varm,KLR,horsch,liu,tj6,plak} and
Kondo-lattice \cite{Prel}. There are rather few works devoted to the
three-band Hubbard model or the Emery model \cite{em,emrei} that is more
realistic for CuO$_2$ planes in high- $T_c$ superconductors (HTSC). For the $
t-J$ model it was shown that the spectral density function $A_h({\bf k}
,\omega )$ of a doped hole revealed a QP peak of intensity $Z_{{\bf k}
}\approx J/t$ and a broad incoherent part that has a width of about $6-7t$.
The QP band bottom corresponds to the momenta ${\bf k}_1=(\pm \pi /2,\pm \pi
/2)$ . Similar results were obtained for the Emery model \cite{Rei,kab}. The
presence of a great incoherent part and small intensity of QP peak indicate,
that bare holes are rather poor elementary excitations even for ${\bf k}$
close to ${\bf k}_1$ .

In order to investigate the hole motion in the $t-J$ model one usually
decouple the hole operator into a spinless fermion and an
antiferromagnetic magnon operator. As a result the zero 
approximation corresponds to the dispersionless band with zero 
energy of the hole. The hopping of the particle appears only due 
to the fermion-magnon scattering which is treated by the usual 
perturbation method in ${\bf k}$-space. For this reason we think 
that in this approach the resulting quasiparticle pole in the 
fermion Green's function takes into account mainly a polaron of a 
large radius. The close situation takes place in the usual treatment of a
hole motion in the effective three-band model \cite{Rei,kab,Oleg1} and the
Kondo-lattice model \cite{Prel}, when one starts from a bare 
hole, not from a magnetic polaron of a small radius.

In the present paper, within the framework of the effective three-band model
we study the spectral density function $A_p({\bf k},\omega )$ of 
a single small polaron,i.e., an excitation which takes into 
account from the very beginning a local hole-spin 
coupling . It is well known that the simplest candidate for such 
a small polaron is an analog of the so-called Zhang-Rice singlet 
in CuO$_4$ plaquette \cite{ZhR},\cite{bmui}. The 
mean-field spectrum $\Omega _{{\bf k}}$ of this excitation is 
well studied \cite{bmui} and it will be used as the zero 
approximation in our treatment. We shall consider the coupling of 
small polaron to spin-wave excitations within the self-consistent 
Born approximation (SCBA) for the corresponding two-time retarded 
Green's function $G({\bf k},\omega )$.

Our motivations to study $A_p({\bf k},\omega )$ and the corresponding
quasiparticle band are the following. First, it 
may be easily shown for the one-hole problem that
 the mean-field energy of the polaron $\Omega _{{\bf k}}$ represents the
center of gravity of the spectral function 

\begin{equation}
\label{cg1}\Omega _k=\int_{-\infty }^\infty \omega A_p(k,\omega 
)d\omega \end{equation}

This means that the minimum of $\Omega _{{\bf k}}$ is the upper bound
 of the actual position for the
quasiparticle band bottom. The SCBA, based on a bare hole Green's 
function, gives the minimum value of quasiparticle energy equal 
to $\omega _{h,\min }=-2.6\tau $ \cite{kab}, for a typical value 
of copper-copper AFM exchange constant $J=0.7\tau $. Here $\tau $ 
is a constant of the effective oxygen-oxygen hopping via an 
intervening copper site (note that our unit of energy is twice 
that of Ref. \cite{kab} $\tau =2t$). As to the value of the small 
polaron mean field band bottom it turns out to be substantially 
lower than $\omega _{h,\min }$, $\Omega _{{\bf k}}=-3.17\tau $, 
for the same value of $J/\tau $. So we may conclude that 
important local correlations are lost in SCBA when we start from 
the bare hole operators.

Secondly, we shall show that a small polaron represents the 
elementary hole excitation much more better than a bare hole 
dressed by magnons within the framework of SCBA. This will be 
manifested by a relatively great intensity of a quasiparticle 
peak in our calculation.

Finally, the mean field spectrum $\Omega _{{\bf k}}$ of the simplest small
spin polaron explicitly depends on the state of 
the antiferromagnetic background . In the case of long range 
order state, $\Omega _{{\bf k}}$ demonstrates a flat band region 
close to the magnetic Brillouin zone boundary \cite{bmui}.  This 
region corresponds to the bottom of the band . Moreover, if one 
takes into account the direct oxygen-oxygen hopping, finite 
temperature and a more complicated form for a small polaron wave 
function, then $\Omega _{{\bf k}}$ reproduces the experimentally 
observed extended saddle point \cite 
{tobin,gofron,abrikos,dessau,king} which is directed along the line $(0,\pi
)-(0,0)$ \cite{barVan}. Therefore it seems important to clear up 
if the quasiparticle band reproduces the peculiarities of $\Omega 
_{{\bf k}}$ dispersion. Below we study this question for 
the simplest variant of the model.

The distinctive feature of our investigation consists in considering the AFM
copper spin subsystem in a spherically symmetric approach \cite{spin,spinb}.
Such an approach is the most appropriate for treating the quantum 2D AFM at
any finite temperature. As a result the scattering of a spin polaron by spin
excitations in the singlet spin background leads to the spectral function
periodicity relative to the full Brillouin zone. Note that the conventional
two-sublattice spin approach leads to periodicity relative to the magnetic
(reduced) Brillouin zone \cite{varm,KLR,horsch,liu,tj6,plak,kab}.

The paper is organized as follows. In Section II we give the derivations for
the self-consistent equation for the Green's function in the case of the
small polaron approach. In Section III we present the procedure that
provides the opportunity to avoid the iterative solution of the
self-consistent equation for complex energies. The procedure is based on the
continued fraction (CF) expansion of Green's function and it 
gives a possibility to calculate consequently the coefficients 
of the CF expansion with the use of the quadrature method.  
In Section IV we 
offer the termination of the continued fraction which leads to 
the correct analytical properties of the resulting Green's 
function.  Numerical results for the self-energies and spectral 
functions, the relation of our results to the previous approaches
 and discussion are given in Section V. Section VI summarizes the 
results. An Appendix 
contains some details of the approach which gives the expression 
for integrals over the spectral density in terms of the chain 
representation of continuous fraction.

Part of our results were presented in Brief Reports \cite{bar1}. The paper
presented here gives further results as well as the description of the new
method and more details about the calculations.

\section{Effective Hamiltonian and Small Polaron Green's function}

Following Refs.\cite{em,emrei,bmui} we adopt the Hamiltonian that
corresponds to one hole problem in the CuO$_2$ plane in HTSC:
\begin{equation}
\label{hamtot}\hat H=\tau \sum_{{\bf r},{\bf a}_1,{\bf a}_2,\sigma ,\sigma
^{\prime }}c_{{\bf r}+{\bf a}_1,\sigma }^{\dagger }c_{{\bf r}+{\bf a}
_2,\sigma ^{\prime }}(\frac 12\delta _{\sigma \sigma ^{\prime }}+2\vec s
_{\sigma \sigma ^{\prime }}\vec S_{{\bf r}})+\frac J2\sum_{{\bf r},{\bf g}}
\vec S_{{\bf r}}\vec S_{{\bf r}+{\bf g}},
\end{equation}
${\bf a}_1,{\bf a}_2=\pm \frac 12{\bf g}_x,\pm \frac 12{\bf g}_y,{\bf g}=\pm
{\bf g}_x,\pm {\bf g}_y.$

Here and below ${\bf g}_{x,y}$ are basic vectors of a copper square lattice
( $|{\bf g}|\equiv 1$), ${\bf r}+{\bf a}$ are four vectors of O sites
nearest to the Cu site ${\bf r}$, the operator $c_\sigma ^{\dagger }$
creates a hole with the spin index $\sigma =\pm 1$ at the O site, $\vec s
_{\sigma \sigma ^{\prime }}=\frac 12\vec \sigma _{\sigma \sigma ^{\prime }}$
, operator ${\bf S}$ represents the localized spin on the copper site. As
mentioned above $\tau $ is the integral of oxygen hole hoppings that takes
into account the coupling of the hole motion with copper spin subsystem. $J$
is the constant of nearest neighbor AFM exchange between the copper spins.

It is well known that the most prominent feature of the Hamiltonian (\ref
{hamtot}) is that the low-energy physics of hole excitations is described by
the Bloch sums ${\cal B}_{{\bf k},\sigma }^{\dagger }$ based on one site
small polaron operators ${\cal B}_{{\bf r}\sigma }^{\dagger }$
\begin{equation}
\label{polar}{\cal B}_{{\bf k},\sigma }^{\dagger }=\frac 1{\sqrt{NK_{{\bf k}
} }}\sum_{{\bf r}}{\rm e}^{i{\bf k\cdot r}}{\cal B}_{{\bf r},\sigma
}^{\dagger },
\end{equation}
\begin{equation}
\label{singl}{\cal B}_{{\bf r},\sigma }^{\dagger }=\frac 12\sum_{{\bf a}}(c_{
{\bf r}+{\bf a},\sigma }^{\dagger }Z_{{\bf r}}^{\bar \sigma \bar \sigma }-c_{
{\bf r}+{\bf a},\bar \sigma }^{\dagger }Z_{{\bf r}}^{\sigma \bar \sigma }),
\end{equation}
$$
K_{{\bf k}}=\langle \frac 1N\sum_{{\bf r},{\bf r}^{\prime }}{\rm e}^{-\imath
{\bf k}({\bf r}-{\bf r}^{\prime })}\{{\cal B}_{{\bf r},\sigma },{\cal B}_{
{\bf r}^{\prime },\sigma }^{\dagger }\}\rangle =1+(C_{{\bf g}}+\frac 14
)\gamma _{{\bf k}}.
$$
Here $\{ ,\} ,[,]$ stand for an anticommutator and commutator, 
respectively; $ \langle ...\rangle \equiv Q^{-1}Sp[{\rm 
e}^{-\beta H}...],\hspace{\parindent} Q=Sp{\rm e}^{-\beta H}$. 
$Sp$ implies taking the trace of an operator, $ \beta =(kT)^{-1}$ 
is an inverse temperature; $\bar \sigma \equiv -\sigma $; $ 
Z_{{\bf r}}^{\sigma _1\sigma _2}\equiv \left| \sigma _1\right\rangle
\left\langle \sigma _2\right| $ are the Hubbard projection operators for Cu
sites states; $\gamma _{{\bf k}}=\frac 14\sum_{{\bf g}}\exp (\imath {\bf
k\cdot g})$.

To calculate the average for commutators and anticommutators such as $K_{
{\bf k}}$, we take into account that these expressions are reduced to the
two-site or three-site averages of the Hubbard operators located 
at different copper sites. In the spherically symmetric approach 
adopted here, all these averages can be expressed in terms of the 
two-site spin correlation functions $C_{{\bf r}}=\langle {\bf 
S}_0\cdot {\bf S}_{{\bf r} }\rangle $. Let us recall that due to 
the spherical symmetry $\left\langle S_i^\alpha S_j^\beta 
\right\rangle =\delta _{\alpha \beta }\left\langle S_i^\alpha 
S_j^\alpha \right\rangle =\frac 13\left\langle {\bf S}_i\cdot 
{\bf S}_j\right\rangle $, $\left\langle S_i^\alpha \right\rangle =0$.
Simultaneously, at $T\rightarrow 0$ the spin subsystem is described by a
long range order state with finite effective magnetization $m$, $C_{{\bf r}
}(|{\bf r|}\rightarrow \infty )=m^2(-1)^{r_x+r_y}$, the value of $m$ is
dictated by the Bose condensation of spin excitations at the
antiferromagnetic vector ${\bf q}_0=(\pi ,\pi )$.

Note, that ${\cal B}_{{\bf k},\sigma }^{\dagger }|L\rangle $ corresponds to
the CuO$_2$ plane state with the total spin equal to $\frac 12$ if $
|L\rangle $ is the singlet state. We treat ${\cal B}_{k,\sigma }^{\dagger }$
as a candidate for the elementary excitations operator and calculate the
corresponding two-time retarded Green's function $G({\bf k},\omega )$ and
spectral density
$$
A_p({\bf k},\omega )=-\frac 1\pi {\rm Im}G({\bf k},\omega +\imath 0^{+}),
$$
\begin{equation}
\label{pGf}G({\bf k},\omega )=\langle {\cal B}_{{\bf k},\sigma }|{\cal B}_{
{\bf k},\sigma }^{\dagger }\rangle _\omega \equiv -\imath \int_{t^{\prime
}}^\infty \!\!dte^{\imath \omega (t-t^{\prime })}\langle \{{\cal B}_{{\bf k}
,\sigma }(t),{\cal B}_{{\bf k},\sigma }^{\dagger }(t^{\prime })\}\rangle .
\end{equation}

The Dyson's equation for $G({\bf k},\omega )$ has the form
\begin{equation}
\label{dy}G^{-1}({\bf k},\omega )=G_0^{-1}-\Sigma ({\bf k},\omega ),\Sigma (
{\bf k},\omega )=\langle {\cal R}_{{\bf k},\sigma }|{\cal R}_{{\bf k},\sigma
}^{\dagger }\rangle ^{(irr)},
\end{equation}
where
\begin{equation}
G_0=(\omega -\Omega _{{\bf k}})^{-1},{\cal R}_{{\bf k},\sigma }=[{\cal B}_{
{\bf k},\sigma },\hat H]=\frac 1{\sqrt{NK_{{\bf k}}}}\sum_{{\bf r}}{\rm e}
^{-\imath {\bf k\cdot r}}{\cal R}_{{\bf r},\sigma },
\end{equation}
\begin{equation}
\label{bdot}{\cal R}_{{\bf r},\sigma }=-4\tau {\cal B}_{{\bf r},\sigma }+
{\cal R}_{{\bf r},\sigma }^\tau +{\cal R}_{{\bf r},\sigma }^J,
\end{equation}
\begin{equation}
{\cal R}_{{\bf r},\sigma }^\tau =-\frac \tau 2\sigma \left( \sum_{{\bf g},
{\bf a},\sigma _1}\sigma _1Z_{{\bf r}}^{\bar \sigma \sigma _1}c_{{\bf r}+
{\bf g}+{\bf a},\bar \sigma _1}-\sum_{{\bf g},{\bf a},\sigma _1,\sigma
_2}\sigma _2Z_{{\bf r}}^{\bar \sigma \sigma _1}Z_{{\bf r}+{\bf g}}^{\sigma
_1\sigma _2}c_{{\bf r}+{\bf g}+{\bf a},\bar \sigma _2}\right) ,
\end{equation}
$$
{\cal R}_{{\bf r},\sigma }^J=\frac J4\sigma \left( \sum_{{\bf g},{\bf a}
,\sigma _1}\sigma _1(Z_{{\bf r}}^{\bar \sigma \sigma _2}Z_{{\bf r}+{\bf g}
}^{\sigma _2\bar \sigma _1}-Z_{{\bf r}+{\bf g}}^{\bar \sigma \sigma _2}Z_{
{\bf r}}^{\sigma _2\bar \sigma _1})c_{{\bf r}+{\bf a},\sigma _1}\right) ,
$$
\begin{equation}
\label{omk}\Omega _{{\bf k}}=\langle \{{\cal R}_{{\bf k},\sigma },{\cal B}_{
{\bf k},\sigma }^{\dagger }\}\rangle =(\tau Q_\tau +JQ_J)/K_{{\bf k}},
\end{equation}
$$
Q_\tau ({\bf k})=-\frac 72-8\left( \frac 14+C_{{\bf g}}\right) \gamma _{{\bf
k}}+\left( \frac 18-C_{{\bf g}}+\frac 12C_{2{\bf g}}\right) \gamma _{2{\bf k}
}+2\left( \frac 18-C_{{\bf g}}+\frac 12C_{{\bf d}}\right) \gamma _{{\bf dk}
},
$$
$$
Q_J({\bf k})=C_{{\bf g}}(\gamma _{{\bf k}}-4),
$$
\begin{equation}
\label{Rirr}\langle {\cal R}|{\cal R}\rangle ^{(irr)}=\langle {\cal R}_{{\bf
k},\sigma }|{\cal R}_{{\bf k},\sigma }^{\dagger }\rangle -\langle {\cal R}_{
{\bf k},\sigma }|{\cal B}_{{\bf k},\sigma }^{\dagger }\rangle \frac 1{
\langle {\cal B}_{{\bf k},\sigma }|{\cal B}_{{\bf k},\sigma }^{\dagger
}\rangle }\langle {\cal B}_{{\bf k},\sigma }|{\cal R}_{{\bf k},\sigma
}^{\dagger }\rangle .
\end{equation}
Here and below ${\bf d}={\bf g}_x+{\bf g}_y,\gamma _{{\bf dk}}=\cos
(k_xg)\cos (k_yg)$.

We see from Eqs.(\ref{dy}) and (\ref{Rirr}), that the self-energy $\Sigma (
{\bf k},\omega )$ accounting for interaction effects is expressed through
the higher-order Green's functions. One should notice, first, that the terms
linear in ${\cal B}_{{\bf k},\sigma }$ do not contribute to the irreducible
Green's function (\ref{Rirr}). Second, the lowest-order self-energy
contribution is provided by the first term in the right-hand side of the
expression (\ref{Rirr}), while the second term leads to higher-order
corrections. Following Ref.\cite{plak} we evaluate (\ref{Rirr}) with a
proper decoupling procedure for two time correlation function $\langle {\cal
R}_{{\bf k},\sigma }(t){\cal R}_{{\bf k},\sigma }^{\dagger }(t^{\prime
})\rangle $. This procedure is equivalent to the self-consistent Born
approximation in a usual diagrammatic technique \cite{plak}. In our case
this means that the two-time correlation function is decoupled into the
spin-spin correlation function and the polaron-polaron correlation function.
The adopted decoupling procedure preserves the main character of polaron
site operator (\ref{singl}) - four hole site operators surround the copper
spin operator. It schematically has the form
$$
\left\langle Z_{{\bf r}_1}(t)\left( \sum_{{\bf a}_1}c_{{\bf r}_2+{\bf a}
_1}(t)Z_{{\bf r}_2}(t)\right) \left( \sum_{{\bf a}_2}Z_{{\bf r}_3}(t^{\prime
})c_{{\bf r}_3+{\bf a}_2}^{\dagger }(t^{\prime })\right) Z_{{\bf r}
_4}(t^{\prime })\right\rangle \simeq
$$
\begin{equation}
\label{decu}\left\langle \left( \sum_{{\bf a}_1}c_{{\bf r}_2+{\bf a}_1}(t)Z_{
{\bf r}_2}(t)\right) \left( \sum_{{\bf a}_2}Z_{{\bf r}_3}(t^{\prime })c_{
{\bf r}_3+{\bf a}_2}^{\dagger }(t^{\prime })\right) \right\rangle
\left\langle Z_{{\bf r}_1}(t)Z_{{\bf r}_4}(t^{\prime })\right\rangle .
\end{equation}
Let us mention that we also tested the more complex decoupling procedure,
and it did not qualitatively alter the results given by approximation (\ref
{decu}). On the next step we project polaron operators in (\ref{decu}) onto $
{\cal B}_{{\bf k}\sigma }$ :
\begin{equation}
\label{proj}c_i(t)Z_j(t)\simeq \xi {\cal B}_{{\bf k}\sigma }(t),\xi
=\left\langle \{c_i(t)Z_j(t),{\cal B}_{{\bf k},\sigma }^{\dagger
}\}\right\rangle .
\end{equation}

Taking into account, that we shall calculate only the irreducible 
part of Green's function (\ref{Rirr}), the averages $\left\langle 
Z_{{\bf r}_1}(t)Z_{ {\bf r}_4}(t^{\prime })\right\rangle $ are 
transformed to corresponding spin-spin correlation functions 
$\left\langle S_{{\bf r}_1}^\alpha (t)S_{ {\bf r}_4}^\alpha 
(t^{\prime })\right\rangle $. Collecting all terms, we have

\begin{equation}
\label{decup}\langle {\cal R}_{{\bf k},\sigma }(t){\cal R}_{{\bf k},\sigma
}^{\dagger }(t^{\prime })\rangle \simeq N^{-1}\sum_{{\bf q}}\frac{K_{{\bf k}
- {\bf q}}}{K_{{\bf k}}}\Gamma ^2({\bf k},{\bf q})\langle {\cal B}_{{\bf k}-
{\bf q},\sigma }(t){\cal B}_{{\bf k}-{\bf q},\sigma }^{\dagger }(t^{\prime
})\rangle \langle {\bf S}_{-{\bf q}}(t){\bf S}_{{\bf q}}(t^{\prime })\rangle
,
\end{equation}
$$
\Gamma ({\bf k},{\bf q})=\tau \Gamma _\tau ({\bf k},{\bf q})+\frac J2\Gamma
_J({\bf k},{\bf q}),
$$
$$
\Gamma _\tau ({\bf k},{\bf q})=4\gamma _{{\bf k}-{\bf q}}\left[ \left(
1+\gamma _{{\bf k}-{\bf q}}\right) /2K_{{\bf k}-{\bf q}}-1\right] ,
$$
$$
\Gamma _J({\bf k},{\bf q})=4\gamma _{{\bf q}}\left[ \left( \frac 34-C_{{\bf
g }}\right) \frac{4\gamma _{{\bf k}-{\bf q}}}{3K_{{\bf k}-{\bf q}}}-1\right]
,
$$
$$
\langle {\bf S}_{-{\bf q}}(t){\bf S}_{{\bf q}}(t^{\prime })\rangle =\frac 1N
\sum_{{\bf r},{\bf r}^{\prime }}{\rm e}^{i{\bf q\cdot }({\bf r}^{\prime }-
{\bf r})}\left\langle {\bf S}_{{\bf r}}(t){\bf S}_{{\bf r}^{\prime
}}(t^{\prime })\right\rangle
$$

Using the spectral representation for Green's functions we obtain the
following intermediate result for the self-energy

$$
\Sigma ({\bf k},\omega )=N^{-1}\sum_{{\bf q}}\frac{K_{{\bf k}-{\bf q}}}{K_{
{\bf k}}}\Gamma ^2({\bf k},{\bf q})\int_{-\infty }^\infty \frac{d\omega _1}
\pi \times
$$
\begin{equation}
\label{fplak}\int_{-\infty }^\infty \frac{d\omega _2}\pi \frac{{\rm e}
^{\beta (\omega _1+\omega _2)}+1}{\left( {\rm e}^{\beta \omega _1}+1\right)
\left( {\rm e}^{\beta \omega _2}-1\right) }\frac{{\rm Im}[G({\bf k}-{\bf q}
,\omega _1+\imath 0^{+})]{\rm Im}[D({\bf q},\omega _2+\imath 0^{+})]}{\omega
-\left( \omega _1+\omega _2\right) +\imath 0^{+}}.
\end{equation}

Here the spin excitation Green's function \cite{spin,spinb}
\begin{equation}
\label{gfspin}D({\bf q},\omega )=\langle S_{-{\bf q}}^z|S_{{\bf q}}^z\rangle
=-\frac{8JC_{{\bf g}}}3\frac{1-\gamma _{{\bf q}}}{\omega ^2-\omega _{{\bf q}
}^2};
\end{equation}
$$
\omega _{{\bf q}}^2=-32J\alpha _1C_{{\bf g}}(1-\gamma _{{\bf q}})(2\Delta
+1+\gamma _{{\bf q}})
$$
we neglect the influence of doped holes on copper spin dynamics and take the
spin spectrum parameters calculated in Ref.\cite{spin} (the vertex
correction $\alpha _1=1.7$ , the spin excitations condensation part $
m^2=0.0225$).

As a result we come to the integral equation for the Green's function that
always arises within the framework of SCBA
\begin{equation}
\label{gineq}G({\bf k},\omega )=\frac 1{\omega -\Omega _{{\bf k}}-\Sigma (
{\bf k},\omega )},
\end{equation}

where
\begin{equation}
\label{gineq1}\Sigma ({\bf k},\omega )=N^{-1}\sum_{{\bf q}}M^2({\bf k},{\bf q
})\left[ (1+\nu _{{\bf q}})G({\bf k}-{\bf q},\omega -\omega _{{\bf q}})+\nu
_{{\bf q}}G({\bf k}-{\bf q},\omega +\omega _{{\bf q}})\right] ,
\end{equation}
$\nu _{{\bf q}}=1/\left[ \exp (\beta \omega _{{\bf q}})-1\right] $ is the
Bose function,
\begin{equation}
\label{vert}M^2({\bf k},{\bf q})=\frac{K_{{\bf k}-{\bf q}}}{K_{{\bf k}}}
\Gamma ^2({\bf k},{\bf q})\frac{\left( -4C_{{\bf g}}\right) \left( 1-\gamma
_{{\bf q}}\right) }{\omega _{{\bf q}}}.
\end{equation}

 $\Gamma ({\bf k},{\bf q})$ corresponds to the bare vertex for 
the coupling between a spin polaron and a spin wave. It is known 
\cite{shrif} that this vertex is substantially renormalized for 
${\bf q}$ close to the AFM vector $ {\bf q}_0=(\pi ,\pi )$ . This 
renormalization is due to the strong interaction of a polaron 
with the condensation part of spin excitations that must be taken 
into account from the very beginning. As a result, the 
renormalized vertex $\tilde \Gamma ({\bf k},{\bf q})$ must be proportional
to \cite{shrif} $\left[ \left( {\bf q}-{\bf q}_0\right) ^2+L_s^{-2}\right]
^{1/2}$, $L_s$ being the spin-spin correlation length, $L_s\rightarrow
\infty $ in our case of a long range order state of the spin subsystem.
Below this renormalization is taken into account empirically by the
substitution
\begin{equation}
\label{renorm}\Gamma ({\bf k},{\bf q})\rightarrow \tilde \Gamma ({\bf k},
{\bf q})=\Gamma ({\bf k},{\bf q})\sqrt{\left( 1+\gamma _{{\bf q}}\right) .}
\end{equation}
The introduced vertex correction is proportional to $\left| {\bf q}-{\bf q}
_0\right| $ for ${\bf q}$ close to ${\bf q}_0$ . We have used also the
following two functions for the vertex correction $\sqrt{1+\gamma _{{\bf q}
}^3}$ and $\sqrt{1+\gamma _{{\bf q}}^5}$ and have obtained the results
similar to those presented below. Let us mention that the bare vertex leads
the dramatic decrease of the QP bandwidth.

\section{Solution of the integral equation}

The equation (\ref{gineq}) is usually solved by an iteration procedure. We
propose here an alternative way, based on the continued fraction 
expansion of $G({\bf k},z)$

\begin{equation}
\label{cf1}G({\bf k},z)=\frac{b_0^2}{z-a_0-}\frac{b_1^2}{z-a_1-}\cdots \frac{
b_n^2}{z-a_n-}\cdots , \quad a_n=a_n({\bf k}),\quad b_n=b_n({\bf k}),
\end{equation}
where
$$
b_0^2=\int_{-\infty }^{+\infty }A_p({\bf k},\omega )d\omega =K_{{\bf k}
},\quad a_0=\frac 1{b_0^2}\int_{-\infty }^{+\infty }\omega A_p({\bf k}
,\omega )d\omega =\Omega _{{\bf k}}.
$$

The coefficients $b_n,a_n,n>0$ are related with the spectral 
density $ A_p({\bf k},\omega )$ via the set of orthogonal 
polynomials $P_n(\omega )$, satisfying the recurrence 
\cite{rm3,rm5,t1,rm7,nex2}: $$ P_{-1}(\omega )=0, \quad  P_0(\omega )=1, $$

\begin{equation}
\label{polrec}P_{n+1}(\omega )=(\omega -a_n)P_n(\omega )-b_n^2P_{n-1}(\omega
),
\end{equation}
and
\begin{equation}
\label{pola}a_n=\frac{\int_{-\infty }^{+\infty }\omega P_n^2(\omega )A_p(
{\bf k},\omega )d\omega }{\int_{-\infty }^{+\infty }P_n^2(\omega )A_p({\bf k}
,\omega )d\omega },
\end{equation}
\begin{equation}
\label{polb}b_{n+1}^2=\frac{\int_{-\infty }^{+\infty }P_{n+1}^2(\omega )A_p(
{\bf k},\omega )d\omega }{\int_{-\infty }^{+\infty }P_n^2(\omega )A_p({\bf k}
,\omega )d\omega }.
\end{equation}

Here we have used the nonnormalized form of the polynomials 
$\int_{-\infty }^{+\infty }P_n(\omega )P_s(\omega )A_p({\bf 
k},\omega )d\omega =\delta _{ns}(\prod_{m=1}^{m=n}b_m)^2$.

Comparing Eqs.(\ref{cf1}) and (\ref{gineq}) we see that the self energy $
\Sigma ({\bf k},z)$ is the continued fraction similar to $G({\bf 
k},z)$ . Thus we can introduce the spectral density

$$
\rho ({\bf k},\omega )=-{\rm Im}[\Sigma ({\bf k},\omega +\imath 0^{+})]/\pi
$$
and the set of polynomials $\Pi _n(\omega )$ with the recurrence analogous
to (\ref{polrec}):
$$
\Pi _n(\omega )=(\omega -a_n)\Pi _{n-1}(\omega )-b_n^2\Pi _{n-2}(\omega
),\quad \quad   \Pi _0(\omega )=1, \quad  \Pi _{-1}(\omega )=0,
$$
where
$$
b_1^2=\int_{-\infty }^\infty \rho ({\bf k},\omega )d\omega ,\quad a_1=\frac 1
{b_1^2}\int_{-\infty }^{+\infty }\omega \rho ({\bf k},\omega )d\omega ,
$$
\begin{equation}
\label{polab}a_{n+1}=\frac{\int_{-\infty }^{+\infty }\omega \Pi _n^2(\omega
)\rho ({\bf k},\omega )d\omega }{\int_{-\infty }^{+\infty }\Pi _n^2(\omega
)\rho ({\bf k},\omega )d\omega },\quad b_{n+1}^2=\frac{\int_{-\infty
}^{+\infty }\Pi _n^2(\omega )\rho ({\bf k},\omega )d\omega }{\int_{-\infty
}^{+\infty }\Pi _{n-1}^2(\omega )\rho ({\bf k},\omega )d\omega },\quad n\geq
1.
\end{equation}

On the other hand, we have from Eq.(\ref{gineq1})
\begin{equation}
\label{roexp}\rho ({\bf k},\omega )=\frac 1{K_{{\bf k}}}N^{-1}\sum_{{\bf q}
}M^2({\bf k},{\bf q})\left[ (1+\nu _{{\bf q}})A_p({\bf k}-{\bf q},z-\omega _{
{\bf q}})+\nu _{{\bf q}}A_p({\bf k}-{\bf q},z+\omega _{{\bf q}})\right] .
\end{equation}

If we shall put the expression for $\rho ({\bf k},\omega )$ in Eq.(\ref
{polab}) then the coefficients $a_{n+1}$ and $b_{n+2}$ will be expressed
through the integrals of the form
\begin{equation}
\label{intw}\int_{-\infty }^\infty (\omega \pm \omega _{{\bf q}})^m\Pi
_i^2A_p({\bf k}-{\bf q},\omega )dw,\quad i\leq n,\quad m=0,1,
\end{equation}

Now, the trick is that the polynomials in $\omega $ in the 
integrals (\ref {intw}) have the degree less than or equal 
$2n+1$. As it was proved by Nex \cite{rm5} such integrals may be 
expressed through the coefficients $ \{a_0,\ldots ,a_n,b_0\ldots 
,b_n\}$.The details of such a procedure are presented in the 
Appendix. So, it turns out that in SCBA we can recursively 
calculate pairs of coefficients $a_{n+1},b_{n+1}$and to obtain $\Sigma ({\bf
k},\omega )$ in the continued fraction form. Of course we must calculate the
coefficients at all the chosen ${\bf k}+{\bf q}$ points in the first
Brillouin zone simultaneously. Below the chosen points correspond 
to a lattice of $32\times 32$ unit cells.The presented procedure 
provides the opportunity to avoid the iterative solution of 
Eq.(\ref{gineq}) for complex energies.

\section{Termination of the continued fraction}

The procedure outlined in the previous section would be efficient if 
after calculating a finite number of coefficients $a_n,b_n,n\leq 
n_0$, we could appropriately approximate the part (infinite 
in our case) of the continued fraction $T_{n_0}$, which has not 
been calculated. In other words, we rewrite the expression 
(\ref{cf1}) in the form \begin{equation} \label{cf2}G({\bf 
k},z)=\frac{b_0^2}{z-a_0-}\frac{b_1^2}{z-a_1-}\cdots \frac{ 
b_{n_0}^2}{z-a_{n_0}-T_{n_0}({\bf k},z)},
\end{equation}
and try to find a function $\tilde T_{n_0}$ (so called ''terminator'') that
is close to $T_{n_0}$.

Various ways to construct such approximations are described in the
literature on the recursion method (see Refs.\cite{rm5,t1,rm7}
). The asymptotic behavior of continued fraction coefficients is governed by
the band structure and singularities of spectral density \cite{rm7}. The
main asymptotic behavior depends on the band structure: $\{a_n\}$ and $
\{b_n\}$ converge towards limits in the single band case, oscillate
endlessly in a predictable way in the multiband case. Damped oscillations
are created by isolated singularities. The main point here is that an
isolated simple pole produces {\em exponentially} damped contribution in $
\{a_n\},\{b_n\}$ , $n\rightarrow \infty $. For our case it means that the
quasiparticle pole position and weight could be obtained with the high
accuracy from finite number of coefficients, and the asymptotic 
behavior determines the incoherent part of the spectrum. It 
is obvious that the spectrum we deal with has an lower bound and 
no upper bound. We can thus expect that coefficients will not 
converge to some finite values but will grow up to infinity.

In Fig.\ref{f1} we represent the coefficients $a_n,b_n$ as functions on $n$
calculated according the procedure described in the preceding section for
two values of $J$, $J=0.7\tau $ and $J=0.1\tau $ and for ${\bf k}=\left( \pi
/2,\pi /2\right) $. As it may be seen, the distinctive feature of this
dependence is that for large $n$ the coefficients $a_n,b_n$ are linear
functions of $n$. For all this, the slope for $a_n$ coefficients is twice as
many as the slope for $b_n$. So the coefficients behavior may be
approximated as

\begin{equation}
\label{asab}b_n\approx \lambda _1n+\lambda _2,\quad a_n\approx 2\lambda
_1n+\lambda _3,\quad \lambda _i=\lambda _i({\bf k}),\quad n\gg 1
\end{equation}
It is interesting that the analogous behavior show the coefficients for $t-J$
model approximated by slave-fermion Hamiltonian \cite{varm} treated within
the self-consistent Born approximation \cite{horsch}. For $J=0.4t,{\bf k}
=\left( \pi /2,\pi /2\right) $ the coefficients $a_n,b_n$ dictated by the
relation analogous to (\ref{gineq1}) are represented in Fig. 
\ref{f2} (a).

Now we shall show that the same asymptotic (\ref{asab}) has the CF expansion
of incomplete Gamma function that is written as \cite{abram}
\begin{equation}
\label{gami1}\Gamma (\alpha ,x)=\frac{{\rm e}^{-x}x^\alpha }{x+}\frac{
1-\alpha }{1+}\frac 1{x+}\frac{2-\alpha }{1+}\cdots
\end{equation}
This circumstance we shall use for the construction of the terminator $
\tilde T_N({\bf k},z)$ for $G({\bf k},z)$ (\ref{cf2})

Let us introduce the function
\begin{equation}
\label{gami2}\tilde g(\alpha ,x)=-\frac{\Gamma (\alpha ,-x)}{{\rm e}
^x(-x)^\alpha }=1/\left( x-\frac{1-\alpha }{1-\theta _1}\right) ,
\end{equation}
where
\begin{equation}
\label{gami3}\theta _n=n/\left( x-\frac{n+1-\alpha }{1-\theta _{n+1}}\right)
.
\end{equation}
In order to rewrite the CF (\ref{gami2}) in the form analogous to Eq.(\ref
{cf1}) we denote
\begin{equation}
\label{gami4}\frac 1{1-\theta _n}\equiv 1+nt_n.
\end{equation}
Then it is not difficult to obtain the relations
\begin{equation}
\label{gami5}t_n=\frac 1{x-(2n+1-\alpha )-(n+1)(n+1-\alpha )t_{n+1}},
\end{equation}
so that
\begin{equation}
\label{gami6}\tilde g(\alpha ,x)=t_0
\end{equation}
has the form (\ref{cf2}) with the coefficients $\tilde b_0^2=1$ and
\begin{equation}
\label{gami7}\tilde a_n=2n+1-\alpha ,\quad \tilde b_n^2=n(n-\alpha ).
\end{equation}
Comparing Eqs. (\ref{asab}) and (\ref{gami7}) for large $n$, when 
$\sqrt{ n(n-\alpha )}\approx n-\alpha /2$, we see that the 
substitution $$ \alpha =-\frac{2\lambda _2}{\lambda 
_1},x=\frac{z+2\lambda _2-\lambda _3+\lambda _1}{\lambda _1} $$ 
leads to the function $\tilde G$
\begin{equation}
\label{gami8}\tilde G({\bf k},z)=\frac 1{\lambda _1}\tilde g(-\frac{2\lambda
_2}{\lambda _1},\frac{z+2\lambda _2-\lambda _3+\lambda _1}{\lambda _1})
\end{equation}
that has the same asymptotic as $G({\bf k},z)$ (\ref{cf1}). It means that $
\tilde G({\bf k},z)$ can be used as the terminator for $G({\bf k},z)$. ,
i.e. we can express $\tilde T_{n_0}({\bf k},z)=\tilde b_{n_0+1}t_{n_0+1}$
through $\tilde G({\bf k},z)$, and the coefficients $\tilde a_n,\tilde b
_n,n\leq n_0$, and then substitute it for $T_{n_0}({\bf k},z)$ (see \cite
{t1} for the details of matching Greenians).

We thus obtain $G({\bf k},z)$ in the whole complex energy plane including
the real axis. Note that usually the procedure of discretizing the energy
range $\omega $ is used for iteration process when the Dyson's equation is
solved numerically. It is not obvious that such a self-consistent solution
leads to the correct analytical properties of the resulting Green's
function. In contrast the continued fraction representation guarantees these
properties (e.g. the positive definiteness of spectral function $A_p({\bf k}
,\omega )$).

\section{Results and discussion}

In this section we represent our results for the retarded Green's function $
G({\bf k},\omega )$ for three-band model at $T=0$. The self consistent
equation (\ref{gineq}) was solved on a $32\times 32$ cell lattice.
The number of calculated CF levels $n_0$ was taken $n_0=30$.

To begin with, we check the validity of the method outlined above 
by calculating the spinless hole Green's function for the $t-J$ 
model and compare the results with the results of Martinez and 
Horsch \cite{horsch} obtained by usual iteration procedure. In 
Fig.\ref{f2} $A_h({\bf k}_1,\omega +\imath \eta ),{\rm Re}\Sigma 
({\bf k}_1,\omega ),-{\rm Im}\Sigma ({\bf k} _1,\omega ),{\bf 
k}_1=(\pi /2,\pi /2)$ for the value of $J=0.4t$ are represented 
for the 16$\times $16 site lattice and broadening constant $\eta 
=0.01t$. The comparison of Fig.\ref{f2} (b),(c),(d) and the 
corresponding functions given by Fig.7,8 from Ref.\cite{horsch} 
(the same lattice size and $\eta $) demonstrates that the 
position of peaks of the hole spectral function and their 
intensities coincide. Some difference is that our $A_h( {\bf 
k}_1,\omega ),$ is more smooth and there are no strong 
oscillations in our self energy $\Sigma ({\bf k},\omega )$ in the 
interval $(-2t<\omega <-0.75t).$

The results for the small spin polaron spectral density, real and imaginary
parts of the self energy for the characteristic value of 
energetic parameter $ J=0.7\tau $ are given in Fig. \ref{f3} 
for the symmetrical points ${\bf k} _1=(\pi /2,\pi /2),{\bf 
k}_2=(0,0),{\bf k}_3=(\pi ,\pi )$. The energy broadening 
parameter is $\eta =0.002$ (we will refer all quantities in units 
of $\tau $ from now on). We find main common feature in the spectral density
for ${\bf k}_1$ and ${\bf k}_2$: a sharp quasiparticle peak exists at the
bottom of each spectrum. The position of the QP peak corresponds to the
condition ${\rm Re}G^{-1}({\bf k},\omega )=0$ , i.e., the point 
where we have the crossing of the functions $y=\omega -\Omega 
_{{\bf k}}$ and ${\rm Re}\Sigma ( {\bf k},\omega )$ , see 
Figs.\ref{f3} (a) and \ref{f3} (b). In Fig. \ref{f3} (d) we 
show $A_p({\bf k}_1,\omega )$ calculated for $\eta =0.002$ (solid 
line) and $\eta =0.0005$ (dashed line) in order to study the 
scaling behavior of the peaks and their widths with respect to 
changes in $\eta $.  Both peaks fit quite closely with a 
Lorentzian $(1/\pi )\left\{ Z({\bf k} _1)\eta /\left[ {\left( 
\omega -\epsilon ({\bf k}_1)\right) }^2+\eta ^2\right] \right\} $ 
, $\epsilon ({\bf k}_1)$ is the location of the peak, which in 
the limit $\eta \rightarrow 0$ becomes $Z({\bf k}_1)\delta 
(\omega -\epsilon ({\bf k}_1))$. It means that ${\rm Im}\Sigma 
({\bf k},\omega _p)\rightarrow 0$ in the same limit. Here and 
below we shall speak about the position $\epsilon ({\bf k})$ of 
such peaks (with imaginary part of pole close to zero) as about 
the energy of the quasiparticle.

Figs. \ref{f3} (a) and \ref{f3} (b) also demonstrate that the 
incoherent part of $A_p({\bf k},\omega )$ increases and the pole 
strength decreases with the increase of $\epsilon ({\bf k})$, 
$Z({\bf k}_1)=0.82,Z({\bf k} _2)=0.347$. Let us recall that 
$\Omega _{{\bf k}}$ represents the center of gravity of the 
spectral function. In our Figures the center of gravity 
corresponds to the crossing of the real axis by the line $y=\omega -\Omega _{
{\bf k}}$. So, if the quasiparticle peak is far from this point 
we would have a great incoherent part.

Quite different features demonstrate $A_p({\bf k}_3,\omega )$ in Fig.\ref{f3}
(c). The broad lowest peak is determined by the appearance of nonzero ${\rm
Im}\Sigma ({\bf k}_3,\omega )$ in the region where ${\rm Re}G^{-1}({\bf k}
_3,\omega )$ has no zeroes. Two broad additional peaks at $\omega 
\approx -1.6$ and $\omega \approx -1.05$ are formed due to the 
zero values of ${\rm Re} G^{-1}({\bf k}_3,\omega )$ close to 
these $\omega $. But at the same time the ${\rm Im}\Sigma ({\bf 
k}_3,\omega )$ is strong in these regions.  Moreover, the maximum 
of ${\rm Im}\Sigma ({\bf k}_3,\omega )$ (near the point $\omega 
\approx -1.37$ ) determines the local minimum of $A_p({\bf k} 
_3,\omega )$ despite of the fact, that this point is close to the 
frequency value where ${\rm Re}G^{-1}({\bf k}_3,\omega )=0$. It 
is clear that it is impossible to treat any of $A_p({\bf 
k}_3,\omega )$ peaks as a quasiparticle one. Let us mind that the 
qualitative behavior of the real part of the self-energy in 
Fig.\ref{f3} (c) is close to that one which is represented by 
Kampf and Schrieffer, see \cite{Kampf} Fig.\ref{f3} (b), for the pseudogap
regime of the Hubbard model. Fig.\ref{f3} (c) demonstrate three solutions of
${\rm Re}G^{-1}({\bf k}_3,\omega )=0$. Although there is a sharp crossover
from a situation with three solutions to one quasiparticle solution, the
spectral function still changes smoothly due to the presence of imaginary
part of $\Sigma $.

Figs. \ref{f3} (b) and (c) demonstrate qualitatively different 
character of $A_p({\bf k},\omega )$ for the points ${\bf 
k}_2=(0,0),{\bf k}_3=(\pi ,\pi )$. This is the consequence of our 
spherically symmetric approach for treating the AFM copper spin 
subsystem. As mentioned in Introduction this approach gives rise 
to the spectral function periodicity relative to the full 
Brillouin zone, not magnetic one.

In Fig.\ref{f4} we show the dispersion relation $\epsilon ({\bf k})$ of the
quasiparticle band and mean field dispersion $\Omega _{{\bf k}}$ along
symmetry lines in the Brillouin zone. For $\epsilon ({\bf k})$ we reproduce
only that ${\bf k}-$ values for which the lowest peak has a pronounced
quasiparticle peak taking the following criteria: -${\rm 
Im}\Sigma ({\bf k} ,\epsilon ({\bf k})+\imath \eta )<2\eta ,\eta 
=0.002$. As it is known\cite {bmui}, due to the antiferromagnetic 
character of spin correlation functions the $\Omega _{{\bf k}}$ 
demonstrates a ''flat dispersion region'' close to the line 
$\gamma _{{\bf k}}<0,|\gamma _{{\bf k}}|\ll 1$ ,i.e. close to the 
boundary of magnetic Brillouin zone X-N-X, see Fig. \ref{f4}. As it may be
seen from Fig. \ref{f4} the quasiparticle band exists for greater part of
the Brillouin zone except the region of the top of the $\Omega _{{\bf k}}$
spectrum. Moreover, the dispersion law $\epsilon ({\bf k})$ qualitatively
reproduces main features of the spectrum $\Omega _{{\bf k}}$. As it is
mentioned in the Introduction, $\Omega _{{\bf k}}$ demonstrates 
the important features of the hole spectrum for CuO$_2$ plane if 
one takes into account O-O hoppings and spin frustration 
\cite{barVan}. We hope that in this case $ \epsilon ({\bf k})$ 
will reproduce these features also.

Let us compare the small polaron spectral density $A_p({\bf k},\omega )$
with the results for the bare hole $A_h({\bf k},\omega )$ given by Kabanov
and Vagov \cite{kab} for ${\bf k}_1=(\pi /2,\pi /2),J=0.7\tau $ , see Fig.
\ref{f3} (a). First, Fig. \ref{f3} (a) indicates that $A_p({\bf k}_1,\omega
) $ demonstrates much sharper QP peaks relative to the results for a bare
hole. For example, the pole strength $Z_p({\bf k}_1)$ for the QP peak of $
A_p({\bf k},\omega )$ equals to $Z_p({\bf k})=0.82$ . The 
corresponding value for $A_h$ given by Ref.\cite{kab} is much 
smaller, $Z_h({\bf k})=0.25$.

Secondly, Fig.\ref{f3} (a) explicitly demonstrates the one peak structure $
A_p({\bf k}_1,\omega )$ in contrast to $A_h({\bf k}_1,\omega )$. Finally, it
is important, that the bottom of our QP $\epsilon ({\bf 
k})=-3.52$ is substantially lower than $\omega _{\min ,h}=-2.6$ 
from \cite{kab}. These results are the consequence of the fact, 
that elementary excitation - spin polaron of small radii ${\cal 
B}_{{\bf k},\sigma }$ - from the beginning takes into account the 
strong local hole-spin coupling.

It is clear that the QP peaks for a bare hole and a small polaron have to
coincide in the exact solution of the problem. The mentioned above
discrepancies between our calculations and that one of Ref.\cite{kab} are
the consequence of different approximations.

In order to test the convergence of our results relative to the increase of
the lattice size and the number of calculated CF levels $NL$, in 
Fig.\ref {f5} we show the QP peak of $A_p({\bf k},\omega )$ at 
${\bf k}=(\pi /2,\pi /2)$ , $J=0.7$ for different lattices and 
$n_0$. This peak, as it is seen from Fig.\ref{f5} (a) and (b), 
changes insignificantly in going from $ 24\times 24$ to $32\times 
32$ cell lattice and from 22 to 30 $NL$.

We consider now the transformation of $G({\bf k},\omega )$ with the decrease
of $J$. In order to clarify how the character of $A_p({\bf k},\omega )$
peaks is changed, in Fig.\ref{f6} we show $A_p({\bf k},\omega )$ 
for the value of $J=0.1$ at points ${\bf k}_1=(\pi /2,\pi 
/2),{\bf k}_2=(0,0),{\bf k} _3=(\pi ,\pi )$. The decrease of $J$ 
leads to the enlargement of broad incoherent part of $A_p({\bf 
k},\omega )$.

As before the flat band region of the QP band bottom enlarges along a
magnetic Brillouin zone boundary. It is represented by the point ${\bf k}_1$
. In Fig.\ref{f6} (a) $A_p({\bf k}_1,\omega )$ demonstrates explicitly a
rather strong QP peak, $Z_p({\bf k}_1)=0.5$ , which corresponds to the 
condition ${\rm Re} G^{-1}({\bf k},\omega )=0$. Quite different 
character of $A_p({\bf k},\omega )$ is typical for ${\bf k}$ that 
correspond to the tops of 
$\Omega _{{\bf k}}$ band: in the low-energy sector for ${\bf 
k}_2,{\bf k}_3$, see Fig.\ref{f6} (b),(c), one observes 
$A_p({\bf k},\omega )$ - peaks with small intensity. For 
example, the pole strength $Z_p$ of such a QP peak for $A_p({\bf k}_2,\omega
)$ equals to $Z_p({\bf k}_2)=0.016$. Taking $\omega _l({\bf k})$ as the
value of $\omega $ corresponding to the center of these lowest by energy
peaks one can see that ${\rm Re}G^{-1}({\bf k},\omega _l({\bf k})\ne 0$ for
the ${\bf k}$ under discussion. Figs. \ref{f6} (b) and (c) demonstrates
 that these
peaks are determined by the peaks in ${\rm Im}\Sigma ({\bf k},\omega )$ at
points $\omega _l({\bf k})$. The self-energy part $\Sigma ({\bf k},\omega )$
occurs through Green's function of a small polaron bounded to a spin waves.
These peaks can be considered as the QP band of such complex states.

If we shall treat the width of the quasiparticle band $W$ as the difference $
\omega _l({\bf k}_2=(0,0))-\omega _l({\bf k}_1=(\pi /2,\pi /2))$ then $W$
turns out to be of order $J$ for small values of $J$ ($J\simeq 0.1$) in line
with the results for the hole Green's function approach \cite{kab}.

It is clear, that for small $J/\tau $ the concept of a small spin polaron
fails  and it is important to estimate the
validity limits of this concept. Our calculations demonstrate, that the
intensity of QP peaks and the structure $A_p({\bf k},\omega )$ do not change
dramatically for ${\bf k}$, corresponding to the band bottom, up to $J/\tau
=0.1$. For example, $Z_p(\frac \pi 2,\frac \pi 2)\approx Z_p(\pi ,0)\approx
0.50$ at $J/\tau =0.1$. So the $J/\tau $ lowest boundary value of the small
spin polaron concept validity is lower than $J/\tau =0.1$.

In Table 1 we give the numerical values $\omega _l({\bf k})$ of the center
position of the lowest $A_p({\bf k}_2,\omega )$ peaks ($\omega _l({\bf k}
)=\epsilon ({\bf k})$ for ${\bf k}$ values where QP peak is observed)
and their pole strength (area under the peak) $Z_p({\bf k})$ for $
{\bf k}=(\pi /2,\pi /2),(0,0),(0,\pi )$ and different values of $J$.

We do not represent the results for large $J$, $J\gg \tau $, as our approach
in the present form fails to describe this limit. Here, from the very
beginning we treat a small polaron by a single site operator $B_{{\bf r}}$ (
\ref{singl}). For large $J$ the mean field static energy of such a state is
proportional to $J$ and such a state is unstable. So in this limit we must
extend the basis of site operators. The simplest way is to include in the
basis the additional operator of a bare hole. In SCBA approximation this
will lead to the system of two selfconsistent equations. As a result all
effects of spin subsystem - hole interaction will be proportional to $\tau
/J $. The more general procedure for the extending of the small polaron
operator basis is outlined in \cite{barVan}.

\section{Summary}

We have studied the small spin polaron motion in the three-band model. The
two-time retarded Green's function was calculated within the framework of
self-consistent Born approximation for $32\times 32$ cell lattice. We have
shown that spin polaron of small radius represents a good approximation to
the true quasiparticle low-energy excitation even at mean-field level. The
account of the self-energy does not crucially change the polaron motion
picture for realistic values of parameters. For quasimomenta ${\bf k}$
values, corresponding to the band bottom, most of the total 
spectral weight is concentrated in the quasiparticle peak (Table 
\ref{t1}). In the same region of ${\bf k}$-space the shape of the 
QP dispersion curve $\epsilon ({\bf k})$ reproduces that of 
mean-field dispersion $\Omega _{{\bf k}}$(Fig.\ref{f4}).

We compare our results with previous investigations\cite{Rei,kab} that
start from the bare hole. We see that the small polaron 
mean-field energy $ \Omega _{{\bf k}}$ lies much lower than the 
QP pole obtained from SCBA for the bare hole. Since $\Omega 
_{{\bf k}}$ determines the center of gravity for the Green's 
function spectral density the actual QP pole position (at least 
for the band bottom) should lie deeper in energy than 
$\Omega _{ {\bf k}}$ (Figs. \ref{f3} and \ref{f6}).This means 
that in the three-band model the important local correlations 
should be taken into account in zero approximation and small spin 
polaron should be constructed. Then the polaron scattering on 
spin waves is of less importance and it may be treated
by perturbation methods.

The conclusion is that the low-energy physics of high-$T_c$ superconductors
should be considered in terms of small spin polaron dynamics. In particular,
that the problem of superconducting hole pairing must be treated as pairing
of these quasiparticles rather than pairing of bare holes.

\section*{ACKNOWLEDGMENTS}

We are grateful to O.A. Starykh and P.Horsch for valuable discussions and
comments. This work was supported, in part, by the INTAS-PFBR organization
under project No. INTAS-RFBR 95-0591, by the Russian Scientific Foundation
for Fundamental Researches (Grant No. 95-02-04239-a), by Russian National
program on Superconductivity (Grant No. 93080).

\section*{Appendix. Chain representation and integrals over the spectral
density}

The integration over spectral density could be done with the quadrature
approach\cite{rm5}. It is very efficient one when applied to the electron
structure calculations \cite{nex2}. Unfortunately, the spectral density we
deal with has no upper bound and exponentially depends on the energy, i.e.
it substantially differs from typical spectral density that appears in a
band-structure calculations. It turned out that the 
direct application of Nex's quadrature approach \cite{rm5} is not 
stable numerically for our purposes. That is why for 
calculations of the integrals (\ref{intw}) we use the chain 
representation of continuous fraction. It means that the CF of 
the form (\ref{cf1}) may be interpreted as the Green's function 
$G(\omega )=\left\langle u_0|(\omega - \hat 
h)^{-1}|u_0\right\rangle $ of the one particle tight-binding 
Hamiltonian $\hat h$ of the semi-infinite one-dimensional 
lattice, the $ a_n,b_n,\left| u_n\right\rangle $ being the site 
energies, nearest neighbor hoppings and on-site basis states 
respectively $$ a_n=\left\langle u_n|\hat h|u_n\right\rangle 
,b_{n+1}=\left\langle u_n|\hat h |u_{n+1}\right\rangle .  $$ We 
introduce the eigenstates $\left| \psi _m\right\rangle $, and 
eigenenergies $E_m$ of the chain Hamiltonian
$$
\hat h=\sum_m\left| \psi _m\right\rangle E_m\left\langle \psi _m\right|
$$
then the Green's function spectral density becomes the local density of
states on the zeroth site of the chain\cite{rm3}
$$
A(\omega )=-\frac 1\pi {\rm Im}G(\omega +\imath 0^{+})=\sum_m\left\langle
u_0|\psi _m\right\rangle \delta (\omega -E_m)\left\langle \psi
_m|u_0\right\rangle .
$$
Then the following identities hold
\begin{equation}
\label{quad}F=\int_{-\infty }^{+\infty }f(\omega )A(\omega )d\omega
=\int_{-\infty }^{+\infty }f(\omega )\sum_m\left\langle u_0|\psi
_m\right\rangle \delta (\omega -E_m)\left\langle \psi _m|u_0\right\rangle
d\omega =
\end{equation}
\begin{equation}
\label{quad2}\sum_m\left\langle u_0|\psi _m\right\rangle f(E_m)\left\langle
\psi _m|u_0\right\rangle =\left\langle u_0|f(\hat h)|u_0\right\rangle .
\end{equation}
Nex has proved\cite{rm5} that for a polynomial $f$ of the degree $2n_0+1$
the integral $F$ for the infinite chain has the same value as the analogous
integral for the truncated chain of the length $n_0+1$. The Hamiltonian of
the truncated chain in the basis of states $\{\left| u_0\right\rangle \ldots
\left| u_{n_0}\right\rangle \}$ has the form of the tridiagonal $
(n_0+1)\times (n_0+1)$ matrix
$$
h_T=\left[
\begin{array}{ccccc}
a_0 & b_1 &  &  &  \\
b_1 & a_1 & b_2 &  &  \\
&  & \cdots &  &  \\
&  &  & a_{n_0-1} & b_{n_0} \\
&  &  & b_{n_0} & a_{n_0}
\end{array}
\right] .
$$

Now, instead of integrating the spectral density function by $\omega$
we directly calculate the matrix 
function $f(h_T)$ in order to take $f_{00}$ matrix element. Then 
$F=f_{00}$ , as it follows from the last identity of Eq.  
(\ref{quad2}). We see that the answer for $F$ is expressed only 
through the first coefficients $\{a_0,\ldots ,a_n,b_0\ldots 
,b_n\}$.


\begin{figure}

\caption{The coefficients $a_n$ and $b_n$ of the continued
fraction expansion of $G_p(k,\omega )$ as functions on $n$ for
${\bf k}=(\pi /2,\pi /2)$: a) $J=0.7\tau $, b) $J=0.1\tau $.
Calculated on the $32\times 32$ cell lattice.  }

\label{f1}

\end{figure}

\begin{figure}
\caption{Results for the hole Green's function $G_h(k,\omega )$ for the
$t-J$ model calculated with the same parameters as in
ref.(\protect\cite{horsch}), $J=0.4t$, ${\bf k}=(\pi /2,\pi /2)$,
$\eta =0.01$, $16\times 16$ site lattice. a) The coefficients $a_n$
and $b_n$ of the continued fraction expansion of $G_h(k,\omega )$
as functions on $n$. b) Spectral function $A_h(k,\omega )$. c)
Real part of the self-energy. d) Imaginary part of the
self-energy.  The unit of energy is $t=1$.}

\label{f2}

\end{figure}

\begin{figure}
\caption{
Spin polaron spectral density $A_p(k,\omega  )$, real and
imaginary parts of the self-energy $\Sigma(k,\omega  )$ calculated
for $J=0.7\tau $ and $32\times 32$ cell lattice and different values of $k$. a)
$k=(\pi /2,\pi /2)$, here we also reproduce the hole spectral
function $A_h(k,\omega )$ obtained in Ref.\protect\cite{kab} and
pointed with h-arrows; b) $k=(0,0)$; c) $k=(\pi ,\pi )$. In Figures
(a-c) $\eta =0.002\tau $, the sloping straight lines represent the
function $\omega - \Omega (k)$. d) The dependence of the
quasiparticle peak of $A_p(k=(\pi /2,\pi /2),\omega )$ for two
values of the broadening factor $\eta $.
The unit of energy is $\tau =1$.}

\label{f3}

\end{figure}

\begin{figure}
\caption{
The dispersion of the QP band $\epsilon (k)$ and the mean field
dispersion  $\Omega _k$ along the symmetry lines in the
Brillouin zone (see inset) for $J=0.7$, $32\times 32$ cell lattice, $\eta
=0.002$. }

\label{f4}

 \end{figure}

\begin{figure}

\caption{
$A_p(k=(\pi /2,\pi /2),\omega)$ for $J=0.7$ and $\eta =0.002$
calculated for a) $NL=30$ and different lattice sizes; b)
$32\times 32$ cell lattice and different numbers of calculated CF
levels $NL$. }

\label{f5}

\end{figure}

\begin{figure}

\caption{
Spin polaron spectral density $A_p(k,\omega )$, real and
imaginary parts of the self-energy $\Sigma(k,\omega )$ calculated
for $J=0.1\tau $, $\eta = 0.002$, $32\times 32$ cell lattice at
three values of $k$:  a) $k=(\pi /2,\pi /2)$, b) $k=(0,0)$, c)
$k=(\pi ,\pi )$.  }

\label{f6}

\end{figure}

\begin{table}
  \caption{Position of the lowest by energy peak  $\omega _l({\bf k})$
and the area under the peak $Z_p({\bf k})$ for different values of $J/\tau $
and {\bf k}}
\label{t1}
\begin{tabular}{cccccccc}
$J/\tau $& $Z_p(0,0)$  & $\omega _l(0,0) $  & $Z_p(\pi /2,\pi /2)$&
 $\omega _l(\pi /2,\pi /2)$  & $Z_p(\pi ,0)$  & $\omega _l(\pi ,0)$  \\
\hline
0.1 & 0.016 & -4.24  & 0.50  & -4.48 & 0.55  & -4.51 \\
0.3 & 0.039 & -3.37  & 0.72  & -4.09 & 0.714 & -4.13 \\
0.5 & 0.174 & -2.714 & 0.793 & -3.79 & 0.738 & -3.83 \\
0.7 & 0.347 & -2.25  & 0.823 & -3.52 & 0.808 & -3.56 \\
\end{tabular}

\end{table}

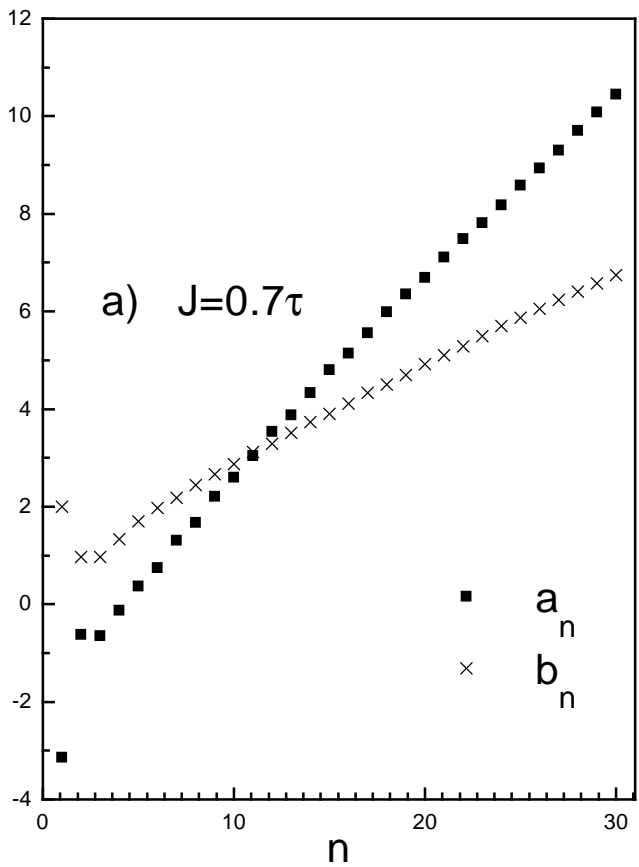
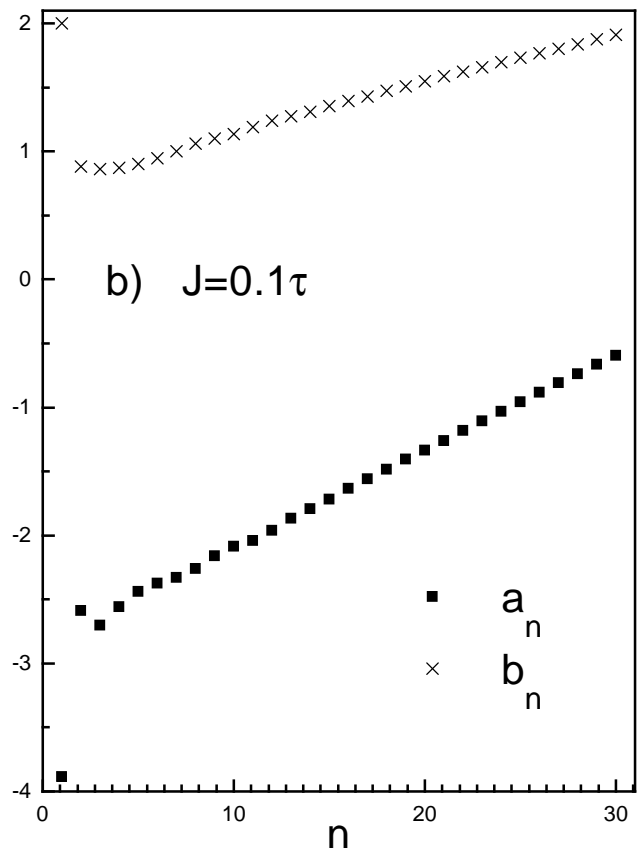
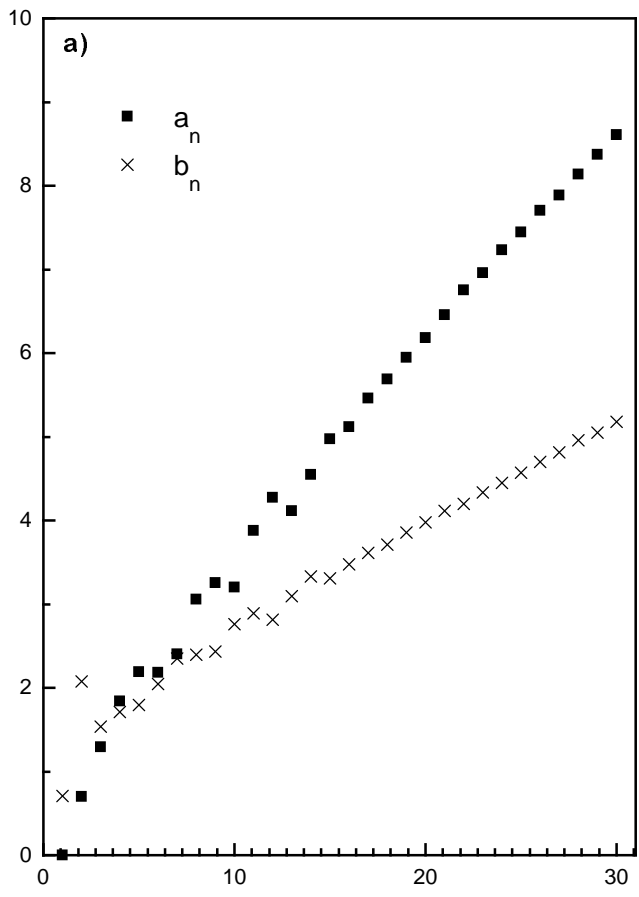
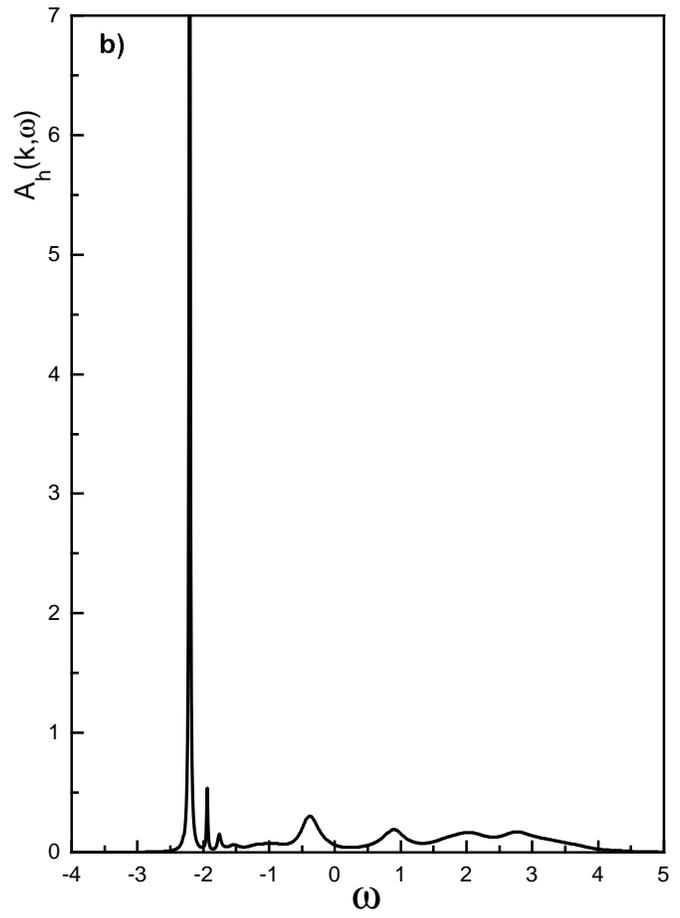
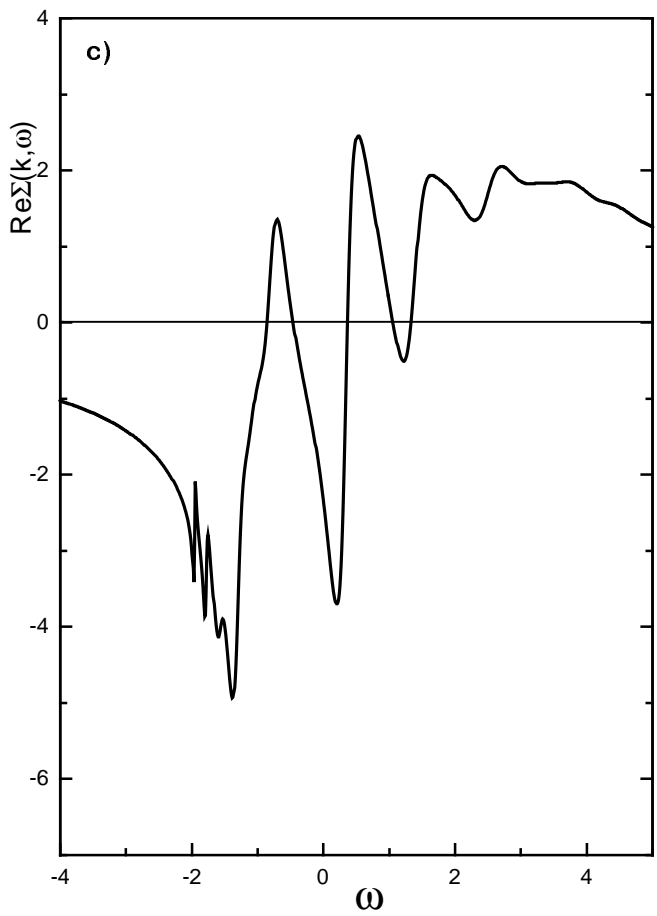
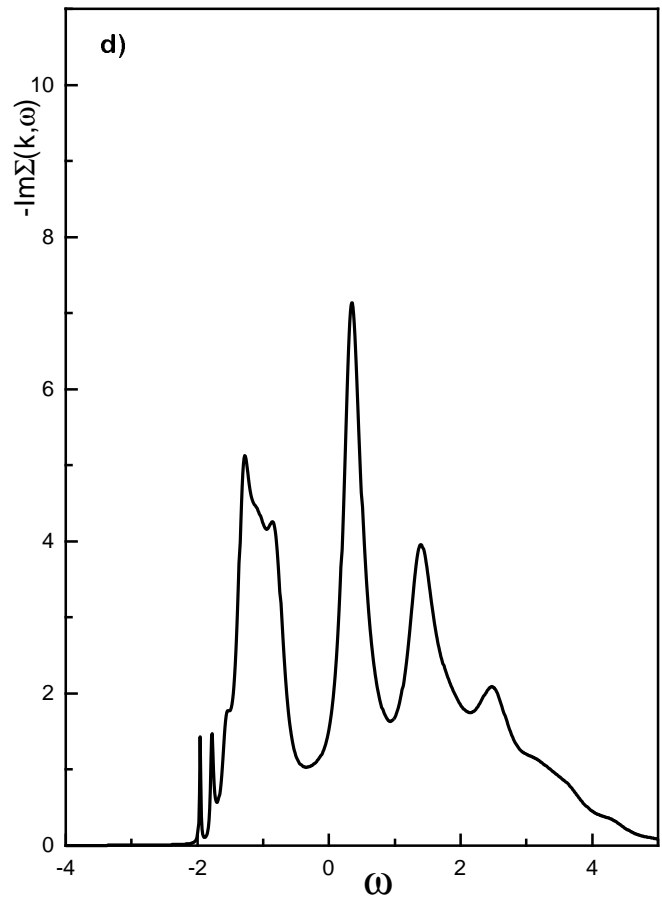
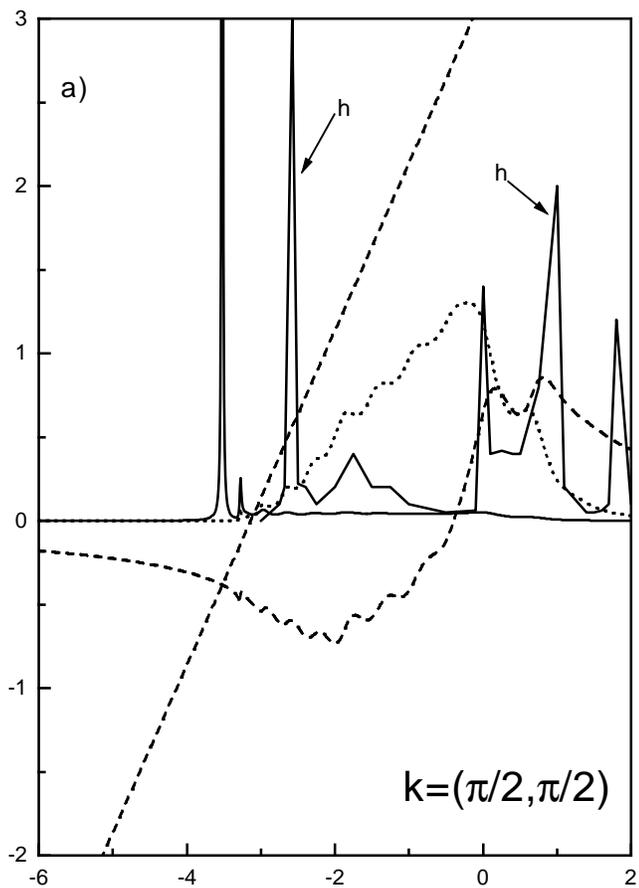
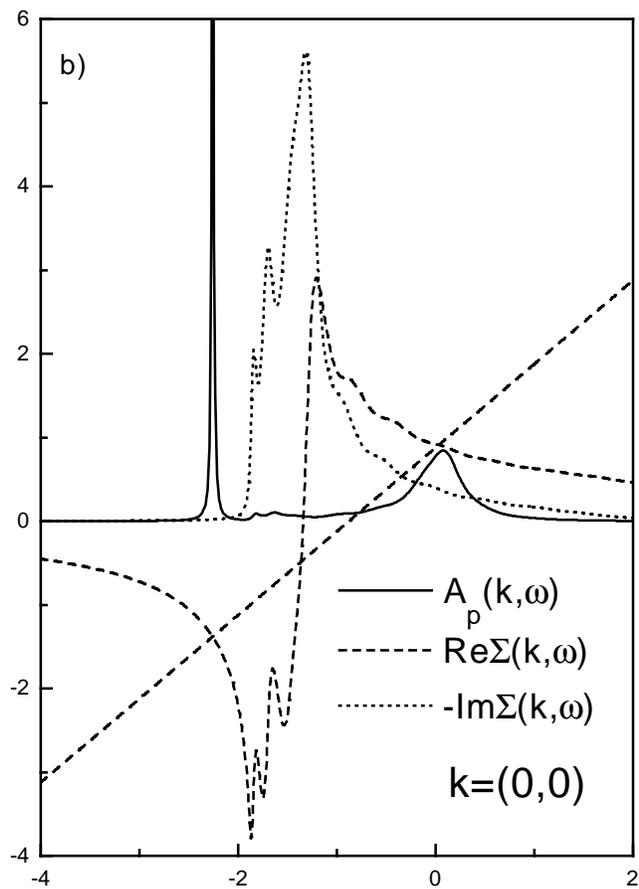
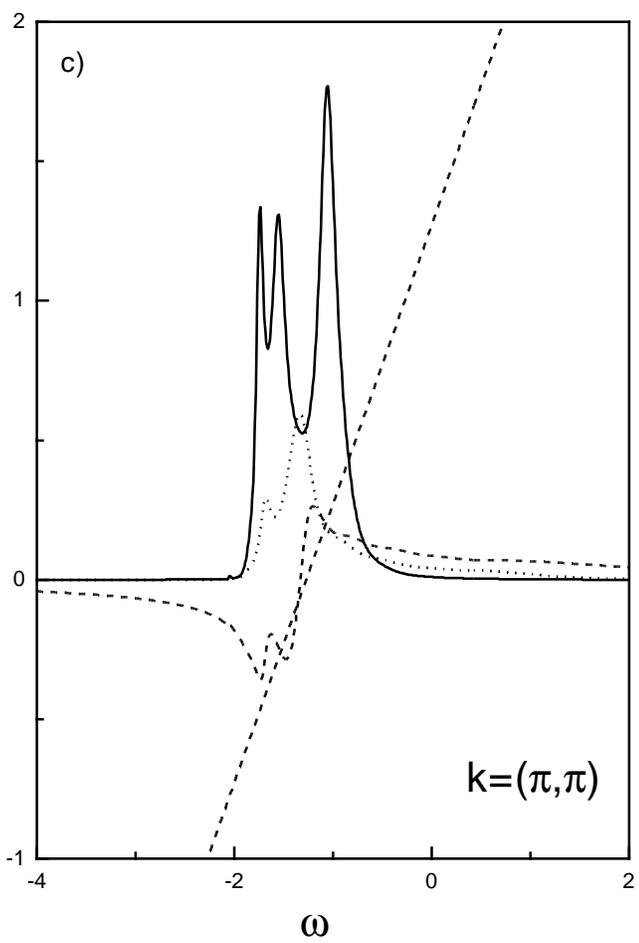
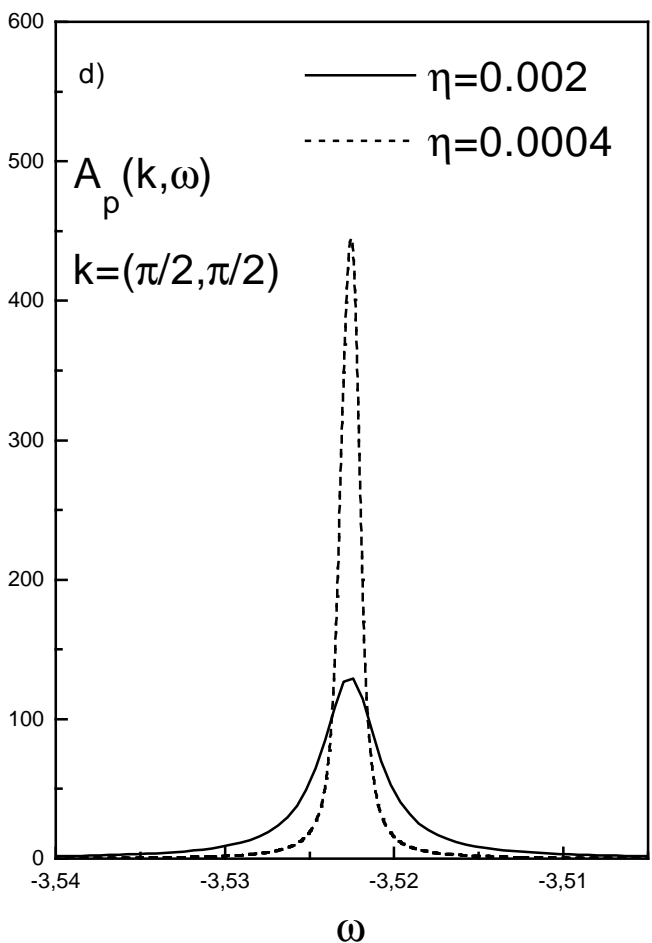
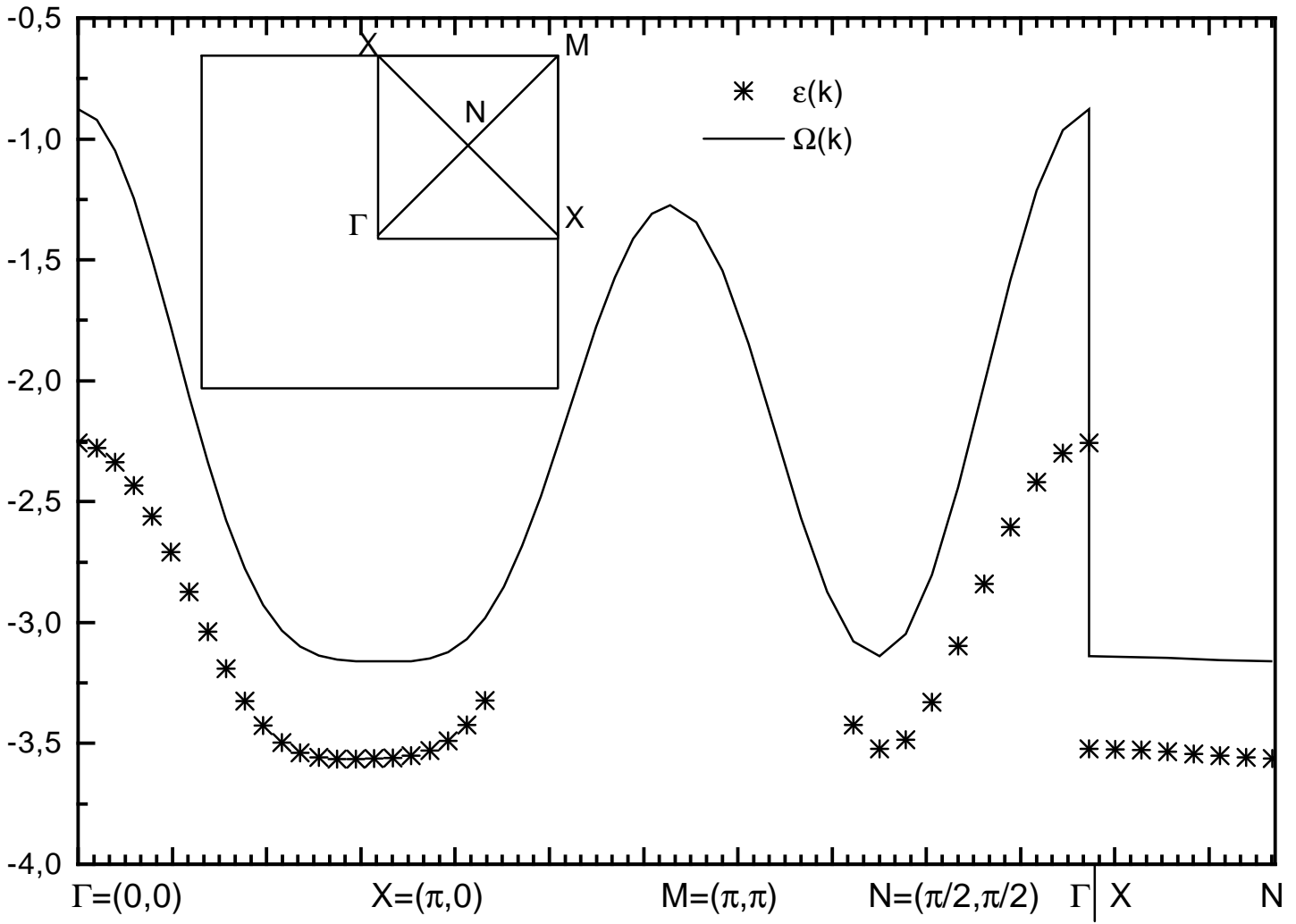
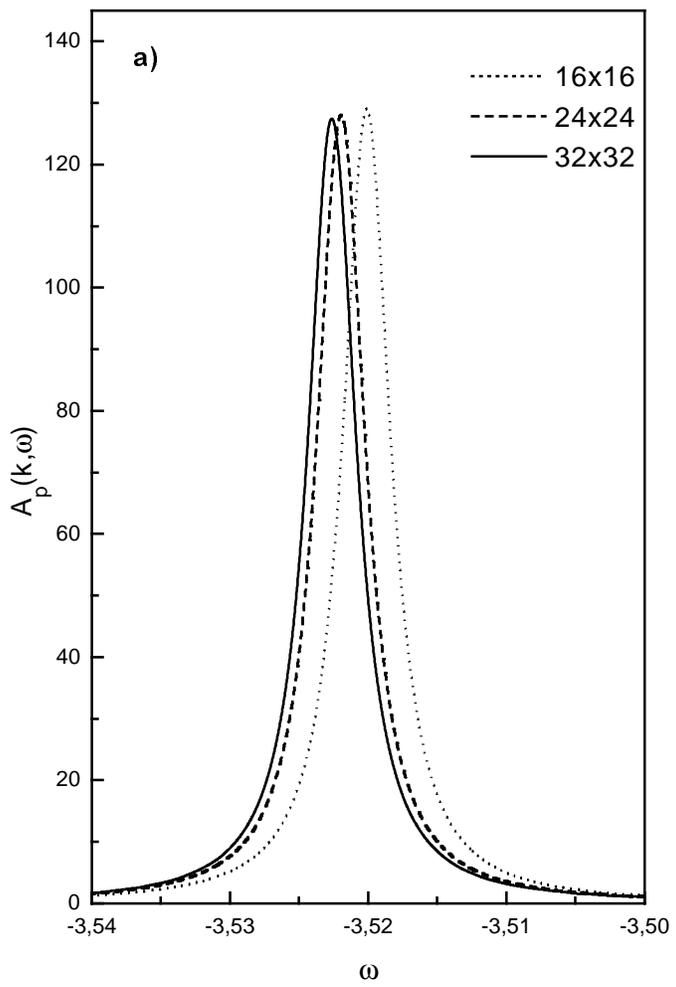
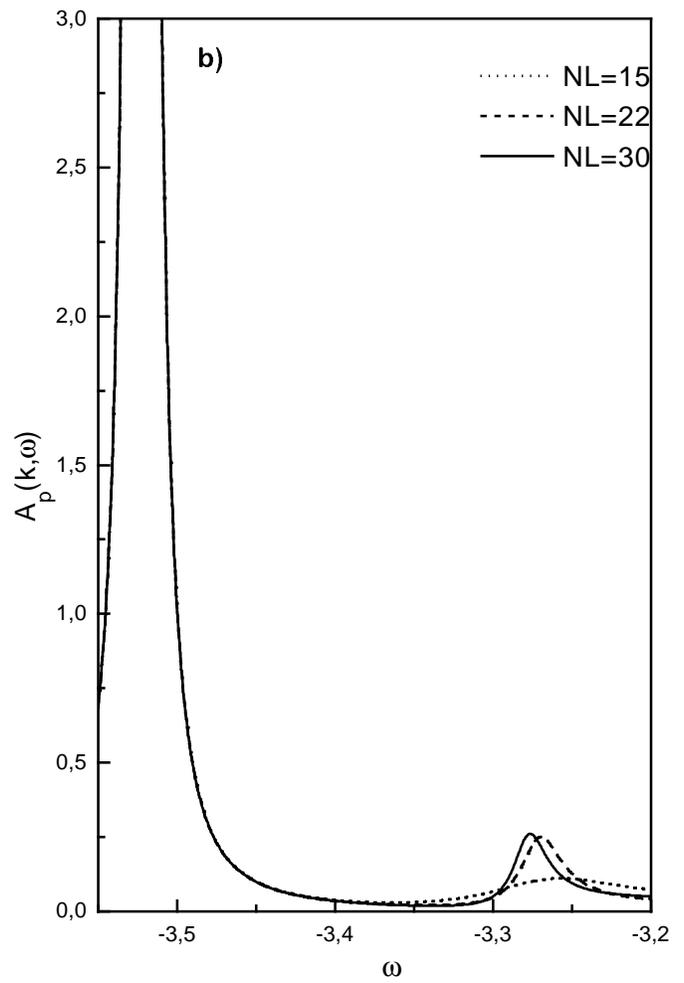
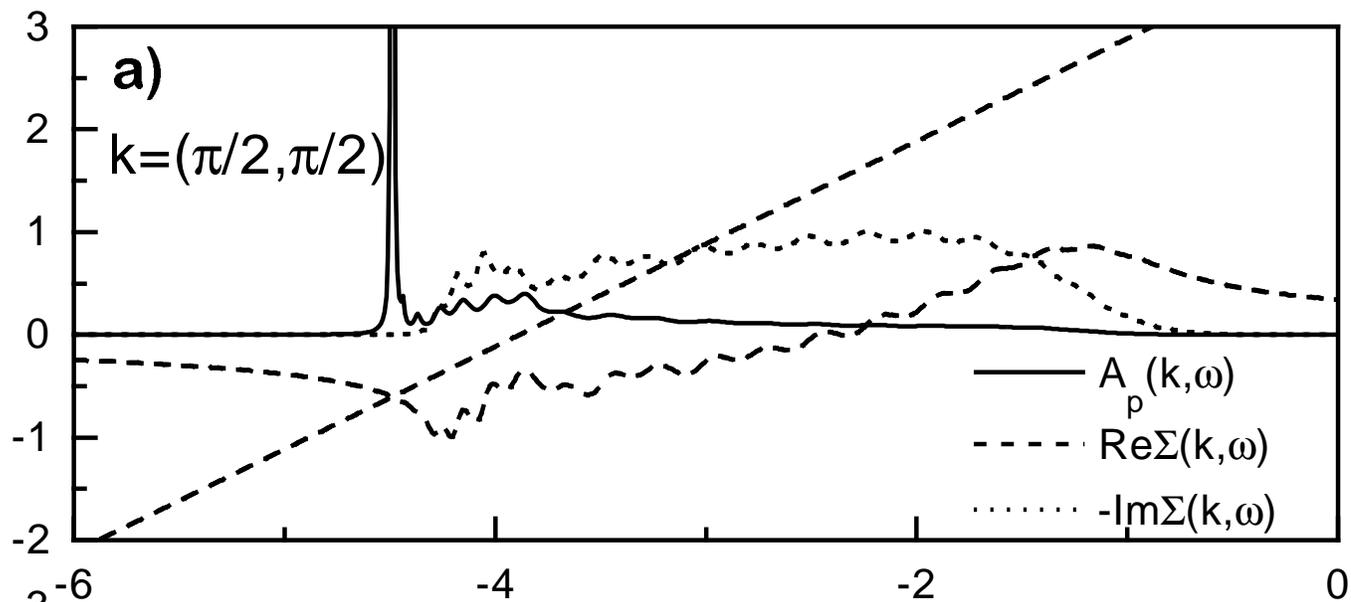
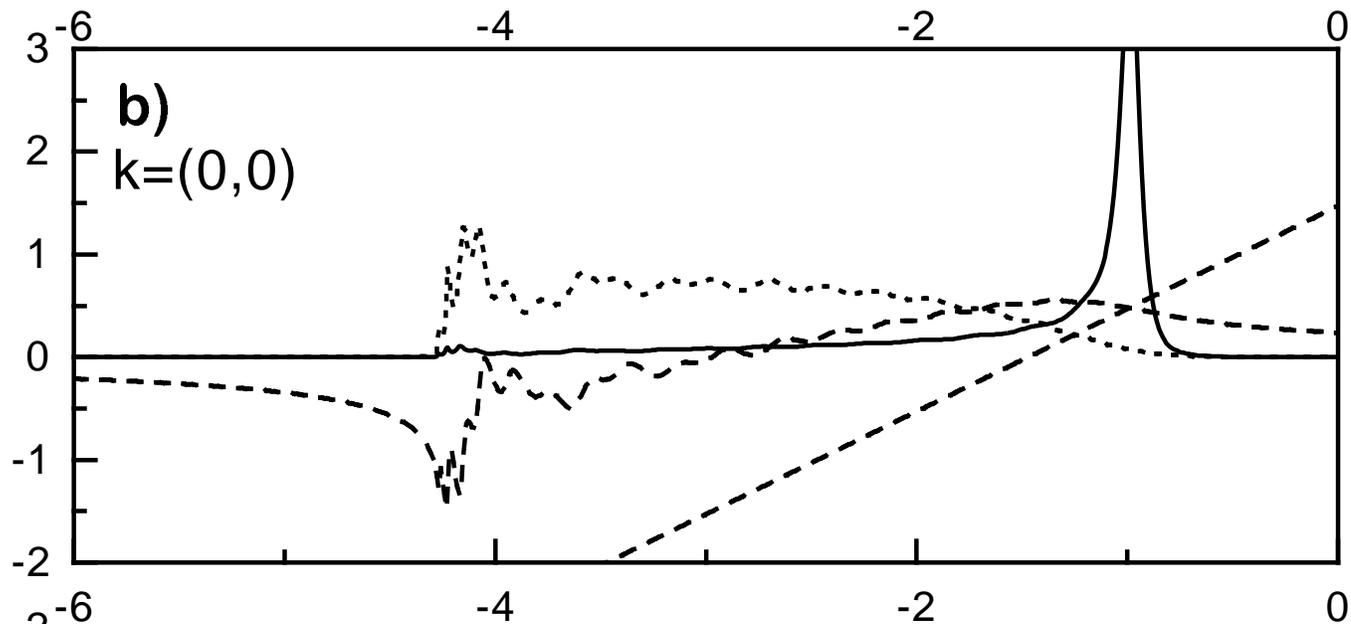
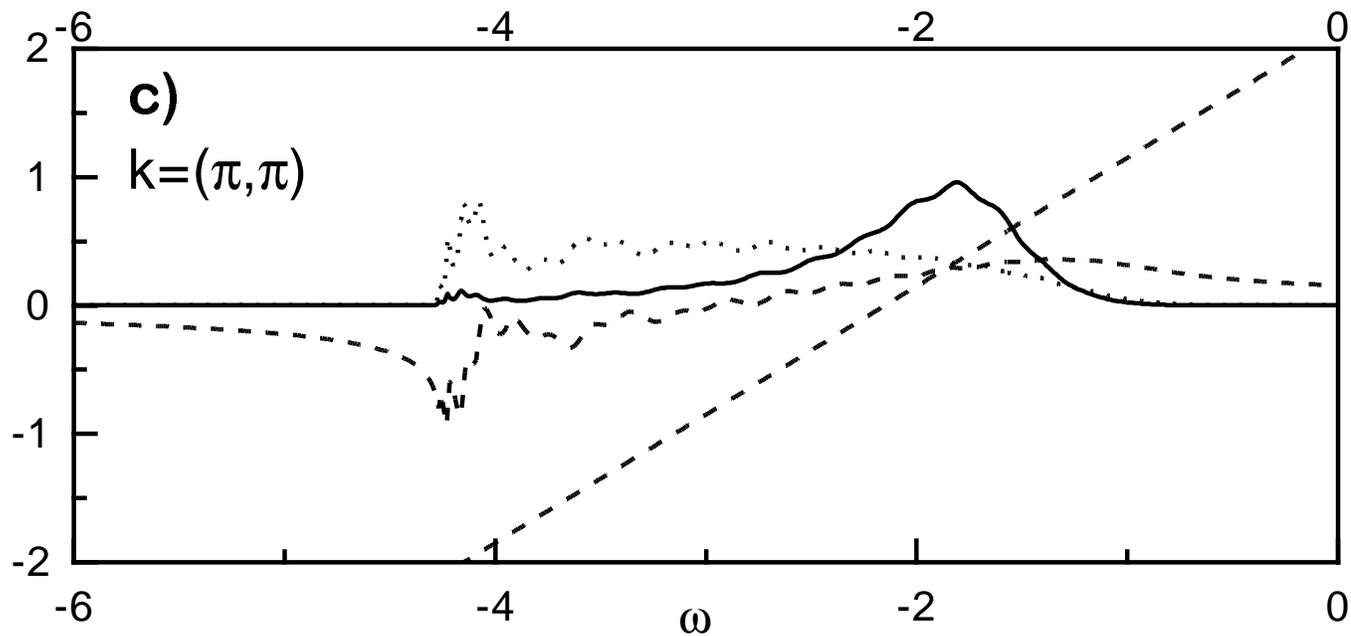
\end{document}